\documentclass[11pt,preprint]{aastex}

\usepackage{lscape}

\begin{document}

\shorttitle{Evolved blue straggler in 47 Tucanae}
\shortauthors{F. R. Ferraro}

\title{Weighing stars: the identification of an Evolved Blue Straggler
  Star in the globular cluster 47 Tucanae\footnotemark[1]}
\footnotetext[1]{Based on UVES-FLAMES observations collected under
  Program 193.D-0232.}

\author{
F. R. Ferraro\altaffilmark{2},
E. Lapenna\altaffilmark{2},
A. Mucciarelli\altaffilmark{2},
B. Lanzoni\altaffilmark{2},
E. Dalessandro\altaffilmark{2},
C. Pallanca\altaffilmark{2},
D. Massari\altaffilmark{3,4}
}
\affil{\altaffilmark{2} Dipartimento di Fisica e Astronomia,
  Universit\`a degli Studi di Bologna, Viale Berti Pichat 6/2,
  I--40127 Bologna, Italy}
\affil{\altaffilmark{3} INAF- Osservatorio Astronomico di Bologna, Via
  Ranzani 1, I--40127 Bologna, Italy}
\affil{\altaffilmark{4} Kapteyn Astronomical Institute, University of
  Groningen, PO Box 800, 9700 AV Groningen, The Netherlands}

\date{Accepted 20 November 2015}

\begin{abstract}
Globular clusters are known to host peculiar objects, named Blue
Straggler Stars (BSSs), significantly heavier than the normal stellar
population. While these stars can be easily identified during their
core hydrogen-burning phase, they are photometrically
indistinguishable from their low-mass sisters in advanced stages of
the subsequent evolution. A clear-cut identification of these objects
would require the direct measurement of the stellar mass.  We used the
detailed comparison between chemical abundances derived from neutral
and from ionized spectral lines as a powerful stellar {\it "weighing
  device"} to measure stellar mass and to identify an evolved BSS in
47 Tucanae.  In particular, high-resolution spectra of three bright
stars located slightly above the level of the "canonical" horizontal
branch sequence in the color-magnitude diagram of 47 Tucanae, have
been obtained with UVES spectrograph.  The measurements of iron and
titanium abundances performed separately from neutral and ionized
lines reveal that two targets have stellar parameters fully consistent
with those expected for low-mass post-horizontal branch objects, while
for the other target the elemental ionization balance is obtained only
by assuming a mass of $\sim 1.4 M_\odot$, which is significantly
larger than the main sequence turn-off mass of the cluster ($\sim 0.85
M_\odot$). The comparison with theoretical stellar tracks suggests
that this is a BSS descendant possibly experiencing its core
helium-burning phase. The large applicability of the proposed method
to most of the globular clusters in our Galaxy opens the possibility
to initiate systematic searches for evolved BSSs, thus giving access
to still unexplored phases of their evolution.
 
\end{abstract}


\section{Introduction}
\label{intro}
Blue straggler stars (BSSs) are commonly defined as stars brighter and
bluer than the main sequence (MS) turn-off point in the
color-magnitude diagram (CMD) of the host stellar cluster. They are
thought to be central hydrogen-burning stars, more massive than the MS
stars \citep{shara97,gilliland98,fiore14}.  Two main processes (able
to bring new hydrogen into the core, thus prolonging the MS stage;
e.g.  \citealp{lom95,lom02,chen09}) are currently favored to account
for their formation: (1) mass-transfer in binary systems \citep{mc64}
possibly up to the complete coalescence of the two stars (see,
however, \citealp{preston00, carney05, gosnell14}), and (2) direct
stellar collisions \citep{hills76}. Both these processes can have an
efficiency which depends on the local environment
\citep{fusi92,ferraro93,ferraro03, davies04, dalessandro08a,
  dalessandro08b, knigge09}, and they can act within the same cluster
simultaneosly \citep{ferraro09,xin15}.  Being more massive than the
average cluster stars, BSSs suffer from the effect of dynamical
friction, that makes them progressively sink towards the cluster
centre \citep{mapelli06, lanzoni07, alessandrini14} and for this
reason they have been found to be powerful probes of the internal
dynamical evolution of the host cluster
\citep[see][]{ferraro12,ferraro15, miocchi15}.

From the stellar evolution side, these objects are expected to
experience all the post-MS evolutionary phases.  However, although
BSSs have been routinely observed since decades now, in globular
clusters, in several open clusters \citep{geller11,gosnell14} and also
in dwarf galaxies \citep{mapelli09,monelli12}, only a few
identifications of evolved BSSs have been obtained so far. Three
candidates have been recently identified in two open clusters (NGC
6791 and NGC 6819; \citealp{brogaard12, corsaro12}) from
asteroseismology studies. Instead only a possible evolved BSS is known
in globular clusters: the anomalous cepheid V19 in NGC 5466
\citep{zinn82,mccharty97} with an estimated mass of $1.6
M_\odot$. Photometric criteria have been suggested to optimize the
search for candidate evolved BSSs: for instance \citet{renzinifusi88}
and \citet{fusi92} suggested to look at a region of the CMD between
the horizontal branch (HB) and the base of the asymptotic giant branch
(AGB, the so-called AGB-clump; \citealp{ferraro99a}), where evolved
BSSs experiencing the core helium burning phase (predicted to be
brighter than {\it canonical}, lower mass, HB stars) are expected to
lie.  Following this prescription, statistical evidence based on
number counts and radial segregation and pointing to the presence of
evolved BSSs "contaminating" the genuine HB-AGB cluster population has
been found in M3 \citep{ferraro97}, M80 \citep{ferraro99b} and 47
Tucanae (47 Tuc; \citealp{beccari06}).

The identification of evolved BSSs is crucial in the context of BSS
formation and evolution models, which currently predict that, at least
at first approximation, these objects experience a post-MS evolution
analogous to that of any ''normal'' star of $\sim 1.2-1.6 M_\odot$
\citep{sills09,tian06}. In fact, the collection of complete samples of
these objects in globular clusters (where BSSs and their descendants
are expected to be numerous enough) allows the determination of
meaningful population ratios from which the characteristic
evolutionary time-scales can be empirically constrained.  Moreover,
since evolved BSSs are 20 times more luminous than their progenitors,
detailed spectroscopic follow-up studies are largely facilitated.  The
determination of the chemical abundance patterns of evolved BSSs is
further helped by the fact that these stars are cooler (thus providing
a larger number of absorption lines) and show slower rotation (hence,
sharper lines), with respect to BSSs on the MS, which are hotter and
often display significant rotation
\citep[see][]{ferraro06,mathieu09,lovisi10,mucciarelli14}.

In this paper we present the results of a high-resolution
spectroscopic exploration of three stars located in a region of the
CMD slightly brighter than the HB red clump in 47 Tuc. Among them, one
object shows chemical properties incompatible with a low-mass object:
the iron and titanium abundance measured from ionized lines agrees
with that obtained from neutral lines only if a mass of $\sim 1.4
M_\odot$ is assumed.  The comparison with stellar tracks suggests that
this objects is probably a BSS descendant experiencing its helium
burning phase.

The paper is organized as follows. The observations, data reduction
and membership determination are discussed in Section \ref{obs}. The
chemical analysis performed on each target and the discussion of its
uncertainties are presented in Sect. \ref{analysis}. Section
\ref{discussion} is devoted to the discussion of the obtained results
in the context of BSS evolution. Sect. \ref{conclusions} summarizes
the results and conclusions of the work.

\section{Observations and membership}
\label{obs}
\begin{figure}[b]
\includegraphics[trim=0cm 0cm 0cm 0cm,clip=true,scale=.70,angle=0]{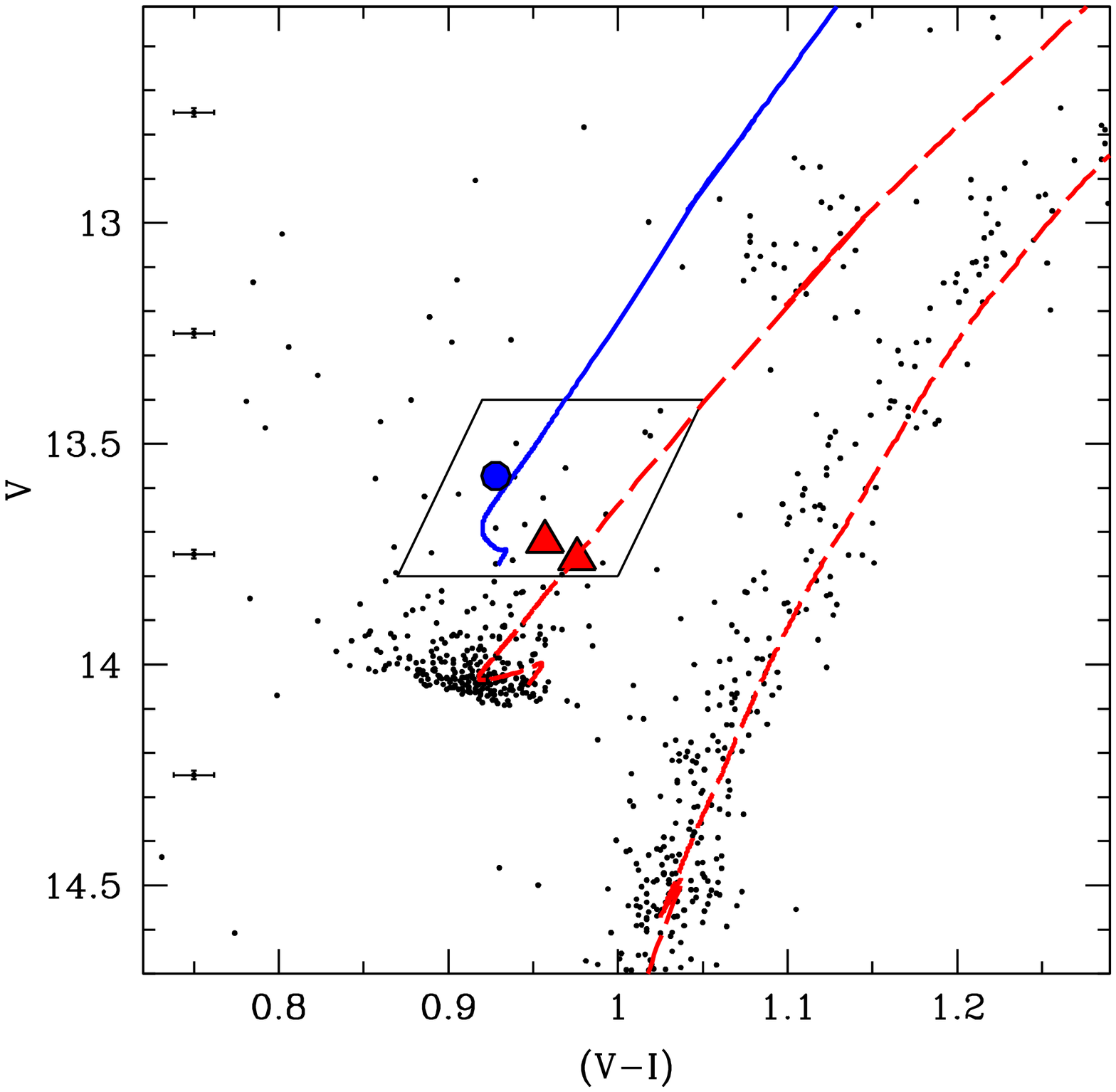}
\caption{Magnified portion of the $(V, V-I)$ CMD of 47 Tuc around the
  horizontal branch.  The large blue circle marks the position of
  E-BSS1. The position of the two reference targets (bHB1 and bHB2)
  are marked with red triangles. For reference, only stars within
  $50\arcsec$ from the center are plotted (small black dots). Error
  bars are marked at different magnitude levels. The red dashed line
  corresponds to the evolutionary track \citep{pietrinferni06} of a
  single star of $0.9 M_\odot$ well reproducing the cluster main
  evolutionary sequences.  The evolutionary track, from the HB to the
  AGB, for a star with a MS mass of 1.5 $M_\odot$ is also shown (blue
  solid line). Because of mass loss during the red giant branch, the
  stellar mass at the HB level for this evolutionary track is $\sim
  1.4 M_\odot$, well in agreement with the spectroscopic estimate for
  E-BSS1. The marked box delimitates the region where evolved BSSs
  experiencing the HB stage are expected to lie: the blue and red
  boundaries of the box are approximately set by the tracks
  corresponding to MS turn-off masses of 1.8 $M_\odot$ and 0.9
  $M_\odot$, respectively.  }
\label{cmd}
\end{figure}

In the context of the ESO Large Programme 193.D-0232 (PI: Ferraro)
aimed at studying the internal kinematics of Galactic globular
clusters, we have secured UVES-FLAMES \citep{pasquini00}
high-resolution spectra of three stars in 47 Tuc.  The targets
(hereafter named bHB1,bHB2 and E-BSS1).  have been selected in a
region of the CMD slightly brighter than the red clump (see Figure
\ref{cmd}), where evolved BSSs experiencing the core helium burning
process are expected to lie (see also \citealp{beccari06}).  All the
targets lie within a distance of $\sim 132\arcsec$ from the cluster
center, corresponding to 4.5 $r_c$ or 0.6 $r_{\rm hm}$
($r_c=29\arcsec$ and $r_{\rm hm}=213\arcsec$ being, respectively, the
core and half-mass radii of 47 Tuc; \citealp{miocchi13}).  Figure
\ref{cmd} shows the ($V, V-I$) CMD obtained from the HST-ACS
photometric catalog of \citet{saraj07}, with the target selection box
marked.  The color and magnitude of bHB2 (which is located beyond the
ACS field of view) are from ground-based wide field data
\citep{ferraro04} homogenized to the Johnson-Cousin photometric
system. The coordinates, $V$ band magnitude, $V-I$ color, and distance
from the center of each target are listed in Table \ref{tab1}.

The target spectra have been acquired with the grating 580 Red Arm
CD\#3, which provides a spectral resolution $R\sim40000$ between
4800$\rm\mathring{A}$\ and 6800$\rm\mathring{A}$. All the spectra have
been reduced by using the dedicated ESO pipeline, performing bias
subtraction, flat-fielding, wavelength calibration and order
merging. During the observations, a number of fibers were allocated on
empty regions to sample the sky background, which has then been
subtracted from the target spectra. The total exposure time is $\sim$
30 min for each star, providing a signal-to-noise ratio (S/N) $\geq$
50 per pixel.

The three stars are all cluster members, as assessed by their radial
velocity (see Table \ref{tab1}) and the systemic velocity and velocity
dispersion of 47 Tuc ($V_r = -17.6$ km s$^{-1}$ and $\sigma$ = 11.5 km
s$^{-1}$, respectively, from \citealp{lapenna14}; see also
\citealp{carretta04, alvesbrito05, ferraro06, koch08, lane10,
  gratton13, cordero14, thygesen14, johnson14}).  The radial
velocities have been determined with the code DAOSPEC
\citep{stetson08}, by measuring the position of up to 300 metallic
lines. The final uncertainty was obtained by dividing the dispersion
of the velocities by the square root of the number of lines used.

Unfortunately no proper motion measures are available for such bright
stars (Andrea Bellini private communication). To further check the
possible contamination by field stars, we extracted the distribution
of radial velocities and metallicities of a sample of about 1700 field
objects from the the Besan\c{c}on Galactic model \citep{robin03}. We
found that no field stars are present in the CMD region corresponding
to the position of the targets and with radial velocities and
metallicities similar to those of 47 Tuc.  We can also safely exclude
a contamination from stars belonging to the Small Magellanic Cloud
(SMC), since the brightest SMC objects (at the red giant branch tip)
are located at much fainter magnitudes ($V \sim 16.5$ mag) and have
quite different radial velocities (between $+50$ and $+250$ km
s$^{-1}$; \citealp{harris06}).

\section{Chemical analysis}
\label{analysis}
The chemical analysis has been performed following the same approach
already used in \citet{lapenna14}. The equivalent widths (EW) and the
relative uncertainty have been measured with DAOSPEC, iteratively
launched by the 4DAO\footnote{http://www.cosmic-lab.eu/4dao/4dao.php}
code \citep{mucciarelli13_4dao}. Abundances were obtained with the
code GALA\footnote{http://www.cosmic-lab.eu/gala/gala.php}
\citep{mucciarelli13_gala}, by matching the measured and theoretical
EWs, and adopting the ATLAS9 model atmospheres and the solar values of
\citet{grevesse98}.  To avoid saturated or too weak features, we have
used only the lines with reduced EWs\footnote{Reduced EWs are defined
  as log($EW/\lambda$), with $\lambda$ being the wavelength.} ranging
between $-5.6$ and $-4.5$ and we have discarded those with EW
uncertainties larger than 20\%. The computation of the final iron
abundances has been performed by using up to 127 FeI lines and 9 FeII
lines. To derive the abundances of titanium we exploited $ 20$ TiI
lines and 11 TiII lines. All the the three targets have been analyzed
using the same line list, which is provided in Table \ref{tab2}.
 
The microturbulence velocity (see Table \ref{tab3}) has been optimized
spectroscopically by requiring that no trends exist between the
abundance derived from FeI lines and the reduced EWs.  To determine
the stellar surface gravity ($\log g$) an estimate of the stellar mass
and radius is needed.  The latter is obtained from the
Stefan-Boltzmann equation once the surface temperature (see Table
\ref{tab1}) has been determined from the $(V-I)_0$ color-temperature
relation of \citet{alonso99}, after transforming the Johnson-Cousin
$(V-I)_0$ color into the Johnson system following appropriate
transformations \citep{bessell79}.  The stellar luminosity has been
evaluated by projecting the CMD position of each star onto the
best-fit isochrone \citep[from][]{pietrinferni06} and assuming a
distance of 4.45 kpc and a color excess $E(B-V)=0.04$
\citep[see][]{harris96}.  \emph{As for the stellar mass, which value
  is expected for the three targets?} Due to the mass-loss occurring
along the red giant branch, stars evolving on the HB are expected to
be less massive than MS turn-off stars by $\sim 0.1-0.15 M_\odot$
\citep{renzinifusi88,origlia07,origlia10,origlia14}.  Recently,
\cite{gratton10} derived the mass distribution of HB stars in several
globular clusters, obtaining values between $\sim 0.6$ and $0.7
M_\odot$ in the case of 47 Tuc. These values are 0.1-0.2 $M_\odot$
lower than the turn-off mass of the best-fit isochrone ($0.85
M_\odot$), in full agreement with the expected amount of mass-loss
during the red giant branch. Adopting a mass of $0.6 M_\odot$ and the
photometric measure of the effective temperature, we obtained surface
gravities of $\sim 2.0$ and we derived the FeI and FeII abundances of
the three targets (see Table \ref{tab3}).
 
For all the targets we found values of [FeI/H] in agreement with the
mean metallicity of 47 Tuc ([Fe/H]$=-0.83$ dex, $\sigma = 0.03$ dex;
\citealp{lapenna14}). This suggests that the target stars are not
affected by departures from local thermal equilibrium
(LTE)\footnote{In fact, departures from LTE conditions affect the
  minority species, leading to an under-estimate of [FeI/H], while
  they have no impact on the abundances obtained from the dominant
  species, as single ionized iron lines \citep[see
    also][]{mashonkina11,ivans01,mucciarelli15}} and the abundance
derived from FeI lines is a reliable measure of the iron content of
the stars. However, only for two objects (namely bHB1 and bHB2) the
value of the iron abundance obtained from the ionized lines (FeII)
agrees, within 0.01 dex, with that derived from FeI, while it is
sensibly ($\sim 0.2$ dex) lower for E-BSS1. This is the opposite of
what is predicted and observed in the case of departures from LTE
conditions, while it could be explained as an effect of surface
gravity and, hence, of stellar mass.  In fact the absorption lines of
ionized elements are sensitive to changes in surface gravity, while
neutral lines are not.  This is due to the fact that, in the case of
highly ionized elements (like Fe), the line-to-continuous opacity
ratio of the neutral lines is independent of pressure (i.e., gravity),
while it shows a dependence in ionized lines
\citep[see][]{gray92}. Taking this into account, we evaluated the
effect of increasing the stellar mass by re-performing the spectral
analysis for different values of the surface gravity.

\begin{figure}[b]
\includegraphics[trim=0cm 0cm 0cm 0cm,clip=true,scale=.80,angle=0]{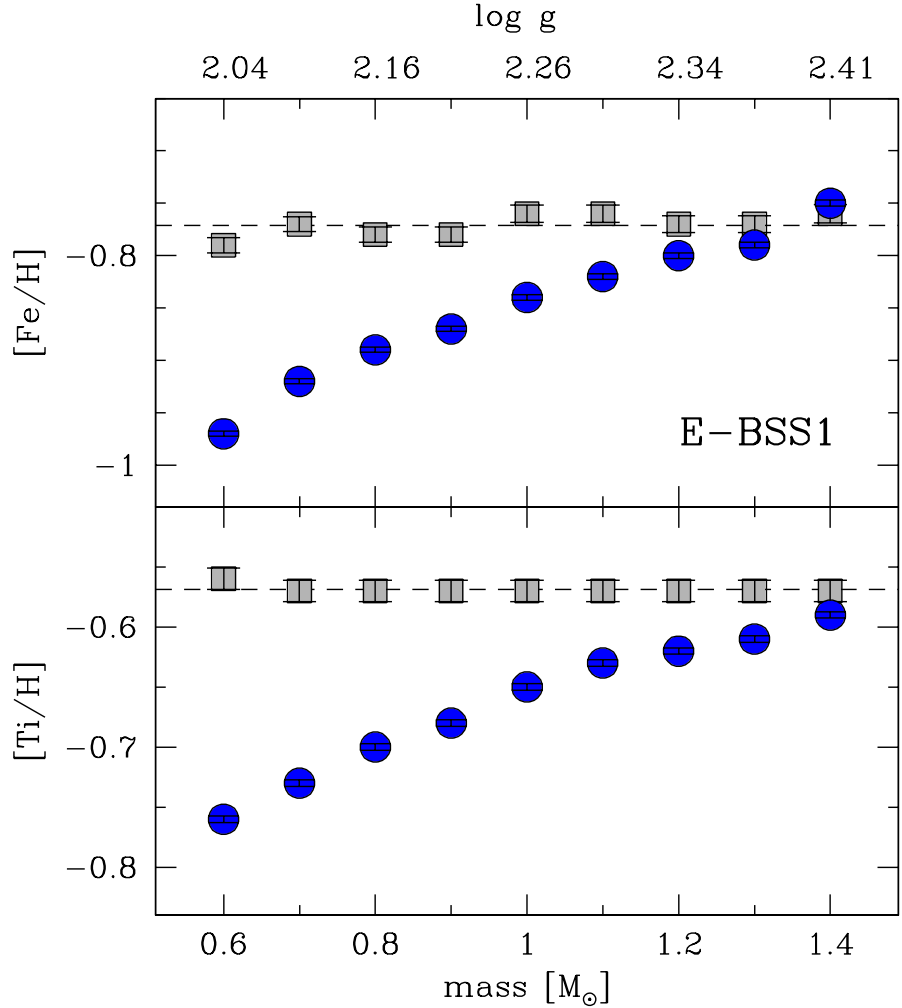}
\caption{\emph{Top panel:} Iron abundance of E-BSS1 derived from FeI
  lines (grey squares) and FeII lines (blue circles), as a function of
  the assumed stellar mass. Errors in the derived abundances are
  smaller than the symbol sizes. The dashed line marks the average FeI
  abundance (well corresponding to the metallicity of 47 Tuc:
  [Fe/H]$=-0.83$ dex; e.g., \citealp{lapenna14}). In the top axis of
  the panel, the logarithmic values of the stellar surface gravity
  corresponding to the various adopted masses are
  labeled. \emph{Bottom panel:} Same as the top panel, but for the Ti
  abundance, as derived from TiI lines (grey squares) and TiII lines
  (blue circles).}
\label{dfeti1}
\end{figure}

\begin{figure}[b]
\includegraphics[trim=0cm 0cm 0cm 0cm,clip=true,scale=.85,angle=0]{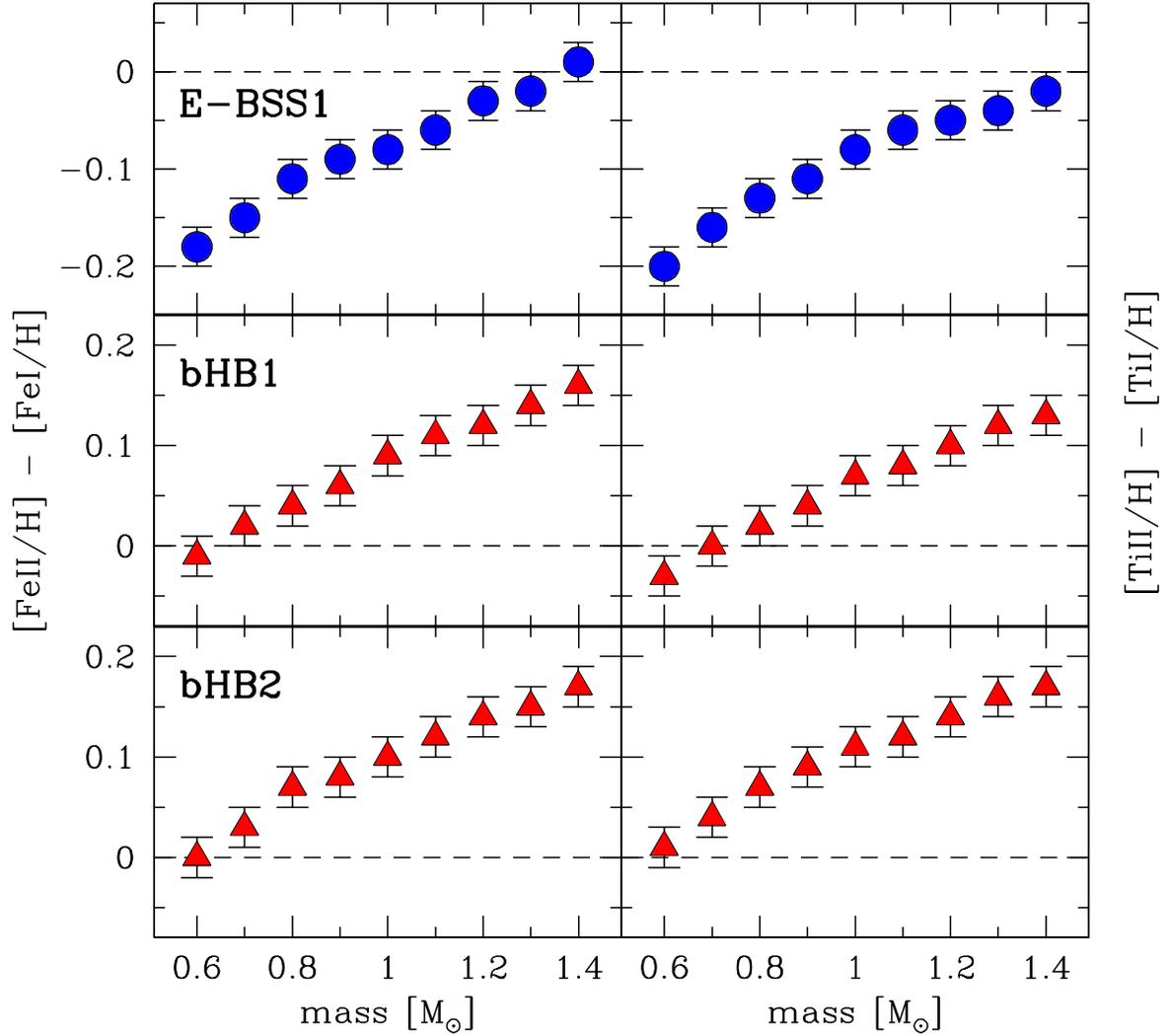}
\caption{\emph{Top panel:} Difference between the iron abundances
  derived from ionized lines and that obtained from neutral lines, as
  a function of the assumed stellar mass (left-hand panel) for E-BSS1.
  The same, but for the titanium abundances is shown in the right-hand
  panels.  \emph{Mid panel:} As in the top panel, but for target bHB1.
  \emph{Bottom panel:} As in the top panel, but for target bHB2.  }
\label{dfeti23}
\end{figure}

In Table \ref{tab3} we list the values of [FeI/H] and [FeII/H]
obtained by varying the star mass in steps of $0.1 M_\odot$ while
keeping the effective temperature fixed. The upper panel of Figure
\ref{dfeti1} shows the resulting behavior. As expected, in all cases
the FeI abundance remains essentially unaltered (and consistent with
the cluster metallicity), while [FeII/H] systematically increases for
increasing mass (gravity).  The behavior of the difference between
FeII and FeI abundances as a function of the adopted stellar mass is
plotted in the left panels of Figure 3 for the three targets.
Clearly, while for two stars (bHB1 and bHB2) a good agreement between
the FeI and FeII abundances is reached at 0.6 $M_\odot$, for E-BSS1 a
significantly larger stellar mass (1.3-1.4 $M_\odot$, corresponding to
a gravity log g = 2.4 dex) is needed. Thus, a mass larger than twice
the mass expected for a canonical post-HB cluster star is needed in
order to reconcile the FeI and FeII abundances of E-BSS1.  Conversely,
the difference [FeII/H]-[FeI/H] for the other two targets tends to
diverge for increasing stellar mass (see Figure \ref{dfeti23}),
indicating that the adopted values of the surface gravity become
progressively unreasonable.

As a double check, the same procedure has been performed on the
titanium lines, since this is one of the few other elements providing
large numbers of both neutral and single ionized lines. Also in this
case, the same abundance of TiI and TiII is reached, within the
errors, only if a mass of 1.4 $M_\odot$ is adopted for E-BSS1, while
the best agreement is reached at 0.6-0.7 $M_\odot$ for bHB1 and bHB2
(see the right panels of Fig. \ref{dfeti23}). This fully confirms the
results obtained from the iron abundance analysis, pointing out that
E-BSS1 is an object significantly more massive than the others.

Additional support comes from the inspection of another feature that
is known to be sensitive to the stellar surface gravity: the wings of
the MgI b triplet at 5167.3, 5172.6 and 5183.6$\rm\mathring{A}$.  In
fact, these lines are located in the damped part of the curve of
growth, where the line broadening, occurring mainly on the wings, is
driven by the collisions between atoms and other particles (the
so-called pressure broadening). In Figure 4 we show a comparison
between the observed spectrum of E-BSS1 and two synthetic spectra
computed by assuming the atmospheric parameters listed in Table
\ref{tab1} and the measured Mg abundance; only the surface gravity has
been varied: we adopted log g=2.03 and log g=2.40 dex (corresponding
to 0.6 and 1.4 $M_\odot$ respectively). Clearly, the synthetic
spectrum computed for log g = 2.03 dex fails to fit the wings of the
MgI b triplet, while that computed assuming a $1.4 M_\odot$ stellar
mass closely reproduces the observed spectrum.  All these findings
point out that E-BSS1 is an object significantly more massive than the
other targets and canonical cluster stars.

\subsection{Uncertainties}
\label{errors}
It is worth noticing that, in doing the comparison between the
abundances derived from the ionized and the neutral species, we
performed a differential analysis, with the most critical parameters
(as temperature, microturbulence and gravity) set to the same
value. Hence only internal errors, due to the quality of the spectra
and the number of the absorption lines used, need to be considered,
while potential external sources of errors can be neglected.

The global uncertainty on the difference between ionized and neutral
chemical abundances has been determined by taking into account the
effect of atmospheric parameter variations and the covariance terms
due to their correlations \citep{cayrel04}. We estimate that the
global effect of varying the temperature by 40 K (corresponding to an
error in color of the order of 0.015 mag) produces a variation of 0.01
dex on the abundance difference. By adding in quadrature this term
with the uncertainties due to the EW measurements (which are of the
order of 0.01 dex for both the abundance ratios), we estimate a total
error of about 0.02 dex on the derived abundance differences.

We emphasize that the only way to make the FeII abundance in agreement
with that of FeI, while simultaneously complying with the other
constraints, is to assume a large mass (gravity) for E-BSS1. In fact,
departures from LTE conditions would affect the neutral species
(yielding to an under-estimate of [FeI/H]), leaving unaltered the
abundances obtained from single ionized lines
\citep{mashonkina11,ivans01,mucciarelli15}. The micro-turbulence
velocity has a negligible impact on the derived abundances, and its
effect is the same (both in terms of absolute value and direction) on
the abundances derived from neutral and from ionized lines: hence, it
cannot help reconciling the value of [FeII/H] with that of
[FeI/H]. Finally, if for E-BSS1 we assume a mass sensibly lower than
1.4 $M_\odot$ (for instance 0.6-0.8 $M_\odot$, as appropriate for HB
and giant stars in 47 Tuc), FeII would agree with FeI only if the
effective temperature is lowered by $\sim 130$ K.  However, this
solution is not acceptable, because it implies a non-zero slope
between the iron abundance and the excitation potential. Moreover,
such a low value of $T_{\rm eff}$ corresponds to a photometric color
that is completely inconsistent with the observed one (note the
internal accuracy of the HST photometry for such a bright object is
less than 0.01 mag). Thus, the spectra inevitably lead to the
conclusion that E-BSS1 is significantly more massive than the other
stars.

\section{Discussion}
\label{discussion}
Indeed the mass derived for E-BSS1 ($1.4 M_\odot$) is by far too large
for a genuine HB cluster star evolving toward the AGB (the MS turn-off
mass in 47 Tuc  is 0.85 $M_\odot$; see
Sect. \ref{analysis}). Moreover, the mass values that we obtained for
targets bHB1 and bHB2 from our analysis turn out to be fully in
agreement with the values (0.6-0.7 $M_\odot$) recently estimated for
typical HB stars in 47 Tuc \citep{gratton10}. Notably, these are also
the values that we obtained for targets bHB1 and bHB2 from our
analysis. These results clearly demonstrate that the detailed
comparison between neutral and ionized chemical abundances is a
powerful {\it weighing device} able to reliably determine stellar
masses in a self-consistent and differential way (this is quite
relevant since it gets rid of any possible zero-point offset among
different methods).

If E-BSS1 were coeval to the other low-mass stars populating the
cluster, such a massive object would have already evolved into a white
dwarf several Gyr ago.  The only possibility is that it formed more
recently, through a mass-enhancement process: it could therefore be
the descendant of a BSS.  As any other star, BSSs are then expected to
evolve along the various post-MS phases. Indeed, E-BSS1 is located in
the region of the CMD where evolved BSSs experiencing the core helium
burning process are expected to lie. In fact, the collisional models
of \citet{sills09} show that, during the core helium burning stage,
the progeny of collisional BSSs should populate the CMD slightly
blueward of the red giant branch, between 0.2 and 1 mag brighter than
the ``canonical'' HB level of the host cluster\footnote{Unfortunately
  no specific tracks for the post-MS evolution of mass-transfer BSSs
  in globular clusters are available at the moment.  However the
  models specifically built for the open cluster M67 \citep{tian06}
  and the globular cluster M30 \citep{xin15} suggest that, after
  mass-transfer, BSSs behave largely as normal single stars of
  comparable mass.  Hence they are also expected to populate the same
  region of the CMD.}.  Overall, the post-MS evolution of a
collisional product in the CMD is very similar to that of a normal
star of the same mass, the former being just a few tens of degree
hotter (see Figure 6 of \citealp{sills09} for the model of a $\sim 1.4
M_\odot$ star).  In the following analysis, we therefore adopted
"normal" evolutionary tracks. In Fig. \ref{cmd} we have superimposed
to the CMD of 47 Tuc the evolutionary track
\citep[from][]{pietrinferni06} of a star with initial mass equal to
$1.5 M_\odot$ (reaching the HB phase with a mass of $1.4 M_\odot$).
We note that E-BSS1 lies in the a region very close to the red clump
level of this track, corresponding to an effective temperature $T_{\rm
  eff} \simeq 5011$ K and a surface gravity $\log g \sim 2.47$ dex.
These values are in very good agreement with those derived from the
chemical analysis of this object.  It seems therefore perfectly
reasonable to identify our star with an evolved BSS that is currently
experiencing its red clump (core helium burning) phase before
ascending the AGB.

{\it How rare evolved BSSs are?} The number of evolved BSSs observable
along the HB stage is predicted to be small even in a massive cluster
like 47 Tuc.  An estimate can be derived by combining the theoretical
ratio between the characteristic MS and HB lifetimes, with the
observed number of BSSs.  In fact, the observational survey of
  \citep{ferraro04} counted 110 BSSs over the entire cluster
  extension, with approximately $40\%$ of the population being
  segregated within 2 core radii ($r<50\arcsec$) from the center,
  which is the distance where E-BSS1 is located.  However, this survey
  sampled only on the brightest portion of the population and deep UV
  observations sampling the entire BSS sequence \citep{ferraro01}
  indicate that the total population is $\sim 1.7$ times larger. Based
  on these numbers, it is reasonable to expect $\sim 75$ BSSs within
  $r<50\arcsec$.
 
On the other hand, evolutionary tracks of collisional BSSs
\citep{sills09} show that the HB lifetime is approximately constant
($\sim 10^8$ years, similar to the HB duration for low-mass single
stars), regardless of the original stellar masses at the time of the
collision.  Instead the MS lifetime of collisional BSSs can change by
2 orders of magnitude.\footnote{Note that, from the comparison between
  Tables 2 and 3 in \citet{sills09}, the MS lifetime of collisional
  products turn out to be $40-70\%$ smaller than that of a normal
  single star with similar mass.} Thus, the predicted number of BSSs
in the HB evolutionary stage sensibly depends on the mass of the BSS
progenitor.  For instance, for a $1.4 M_\odot$ BSS originated from the
collision of a $0.8+0.6 M_\odot$ pair, \citet{sills09} predict $t_{\rm
  MS}=0.82$ Gyr and $t_{\rm HB}=0.096$ Gyr, thus yielding $t_{\rm
  MS}/t_{\rm HB}=8.5$, while this ratios turns out to be 17.7 on
average.  By considering these two values as reasonable extremes for
the ratio $t_{\rm MS}/t_{\rm HB}$, we expect to observe 4-9 evolved
BSSs experiencing the helium burning phase out of a total population
of 75 BSSs (in the MS stage).

Note that in the same region of the cluster ($r<50\arcsec$), several
hundreds genuine low-mass stars, with quite similar photometric
properties and experiencing the same evolutionary phase, are
observed. However, because of their larger mass, evolved BSSs in the
HB stage are expected to appear brighter than "normal" low-mass HB
stars and to lie along the path of low-mass stars evolving toward the
AGB. For genuine low-mass stars the transition from the HB to the AGB
phase (see the box in Figure 1) is quite rapid: $\sim 3.5$ million
years, corresponding to approximately $4\%$ of the time they spent in
the HB phase.  Based on these considerations and the fact that within
$50\arcsec$ from the cluster center we count 270 objects in the HB
clump, we would have expected to observe $\sim 11$ stars within the
box drawn in Figure 1. Instead 20 stars are counted. According to
these estimates, roughly half of the stars within $50\arcsec$ from the
cluster center and lying in the selection box could be evolved BSSs.
Hence beside E-BSS1, 8 additional stars in the box could be evolved
BSSs. This is in very good agreement with the prediction above, based
on the number of BSSs observed on the MS. It is also consistent with
previous evidence \citep{beccari06} showing that the radial
distribution of supra-HB stars in 47 Tuc is anomalously segregated in
the center, as expected if a significant fraction of them is made of
objects heavier than the average, sunk to the bottom of the potential
well because of the cluster dynamical evolution \citep{ferraro12}.

\section{Summary and conclusions}
\label{conclusions}
In this work we have performed the chemical analysis of three stars
observed between the HB and the AGB regions in the CMD of the Galactic
globular cluster 47 Tuc.  By using high-resolution spectra acquired at
the Very Large Telescope, we have used the difference between iron and
titanium abundances derived from neutral and ionized lines as a {\it
  weighing device} to derive the stellar mass.  This provided
convincing evidence that one target (E-BSS1) is significantly more
massive ($\sim 1.4 M_\odot$) than normal cluster stars, while the
other two targets (bHB1 and bHB2) have masses of 0.6-0.7 $M_\odot$,
perfectly consistent with the theoretical expectations. These results
clearly demonstrate that the detailed comparison between neutral and
ionized chemical abundances is a powerful {\it weighing device} able
to reliably determine stellar masses in a self-consistent and
differential way. The presence of such a high-mass star in that region
of the CMD strongly suggests that it is an evolved descendants of a
BSS, caught during its core He-burning phase. Interestingly, the ratio
between the characteristic MS and HB evolutionary times and the number
of BSSs observed in 47 Tuc suggest that a few other evolved BSSs
should populate the same region of the CMD.

The proposed {\it weighing device}, by efficiently pinpointing (heavy)
evolved BSSs into the dominant and photometrically indistinguishable
population of genuine (low-mass) stars, opens the route to systematic
searches of evolved BSSs in globular clusters, thus allowing detailed
spectroscopic follow-up studies with the concrete possibility to even
go back to the formation channel. In fact, a few characterizing
features impressed by the formation process could be still observable
in such advanced stages of the evolution. One of the most solid
predictions of the mass-transfer scenario is that mass-transfer BSSs
should be bound in a binary system with a compact (degenerate)
companion star (the peeled core of the donor, probably a helium white
dwarf). This was recently confirmed in the open cluster NGC 188
\citep{gosnell14}. Thanks to the high luminosity of evolved BSSs,
spectroscopic follow-up observations would make such a prediction
easily testable through the measurement of periodic radial velocity
variations. Since no companion is expected in the collisional scenario
(which ends up with the merger of the two progenitors), this kind of
studies is particularly important in globular clusters, where both
formation channels are expected to be active but their relative
efficiency is still unknown. Moreover, detailed spectroscopic
follow-ups providing the entire chemical pattern of evolved BSSs
represent an additional and highly fruitful route toward the full
characterization of their evolution. In fact, significant depletion of
chemical species like carbon and oxygen has been observed in a
sub-sample of BSSs in 47 Tuc and it has been interpreted as the
chemical signature of the mass-transfer origin of these objects
\citep{ferraro06}. However it is still unknown whether this signature
is transient and on which time-scales. Hence, any additional
information obtained from more advanced stages of the evolution can
provide new clues about the degree of mixing experienced by these
stars. Indeed, after the detection of E-BSS1, the proposed {\it
  weighing device} promises to boost the identifications of evolved
BSSs, thus providing unprecedented constraints to the theoretical
modeling of these exotica and opening a new perspective on the
comprehension of their evolutionary paths and formation processes.

\acknowledgements This paper is dedicated to Prof. Flavio Fusi Pecci,
one of the first scientists indicating the route to search for evolved
BSSs in GCs: his pioneering work inspired our research over the years.
We thank the referees for useful comments that improved the paper
presentation. This research is part of the project {\it Cosmic-Lab}
(see http://www.cosmic-lab.eu) funded by the European Research Council
(under contract ERC-2010-AdG-267675).  E.L.  acknowledges the {\it
  Marco Polo} program for a grant support and the Institute for
Astronomy, University of Edinburgh \& STFC (U.K.), for the hospitality
during the period when most of this work was carried out.


\begin{figure}[b]
\includegraphics[trim=0cm 0cm 0cm 0cm,clip=true,scale=.80,angle=0]{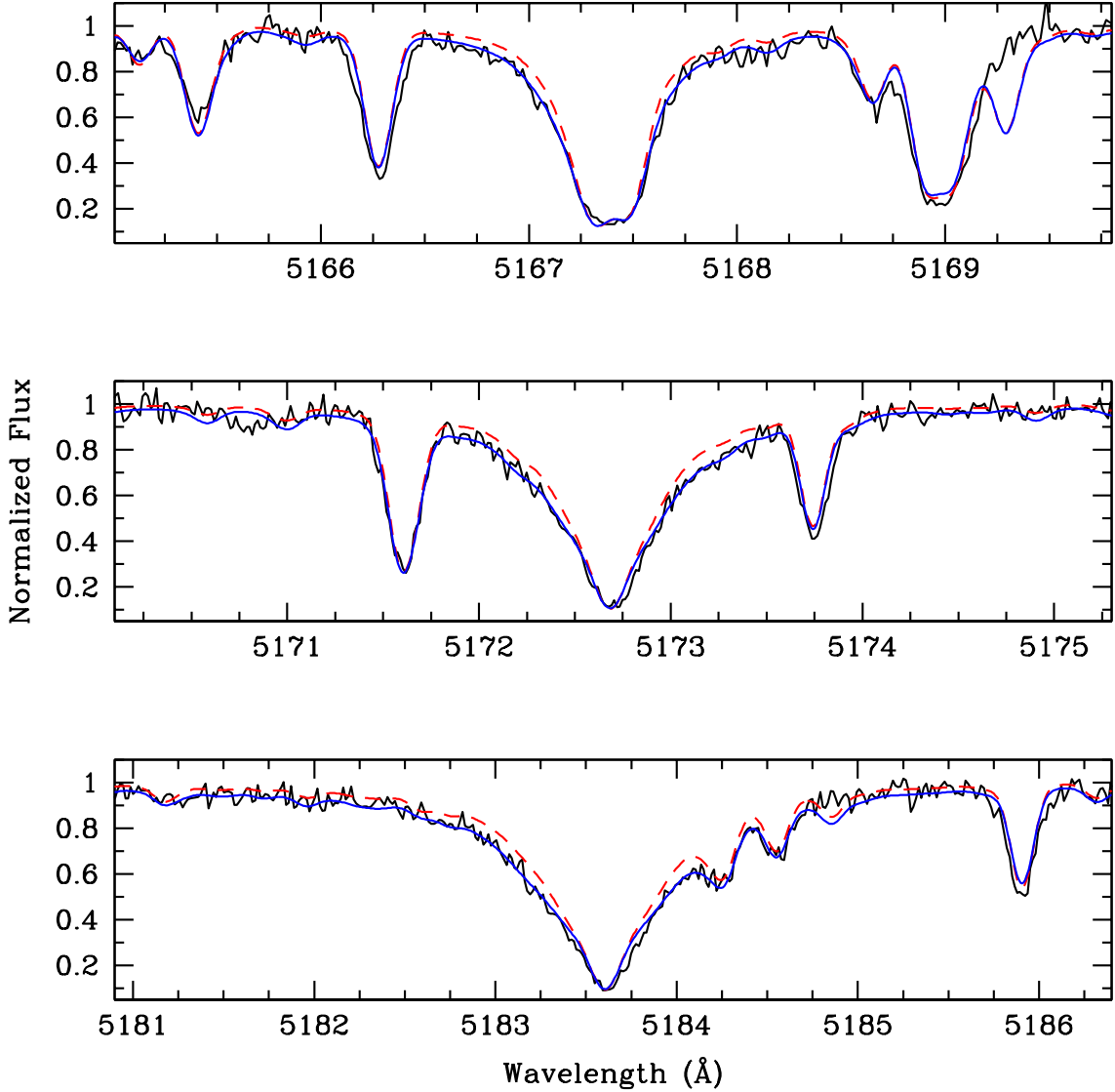}
\caption{Comparison between the observed spectrum (solid black line)
  and two synthetic spectra for the MgI lines at 5167.3, 5172.6 and
  5183.6$\rm\mathring{A}$. The synthetic spectra have been computed by
  adopting $T_{\rm eff} = 5013$ K and $v_{\rm turb} = 1.20$ km
  s$^{-1}$ and by adopting two different values of the surface
  gravity: $\log g = 2.03$ dex corresponding to a stellar mass of $0.6
  M_\odot$ (dashed red line) and $\log g = 2.40$ dex corresponding to $1.4
  M_\odot$ (solid blue line). Clearly, the spectrum obtained for a
  $0.6 M_\odot$ star is unable to reproduce the observations, while
  the observed wings of the MgI b triplet are very well matched under
  the assumption of a $1.4 M_\odot$ stellar mass.}
\label{mglines}
\end{figure}


\begin{deluxetable}{cccccccrr}
\tablecolumns{9}
\tiny
\tablewidth{0pt}
\tablecaption{Observational parameters of the three targets}
\tablehead{\colhead{Name} & \colhead{ID} & \colhead{R.A.}
& \colhead{Dec} & \colhead{$V$} & \colhead{$(V-I)$} & \colhead{$r$} & \colhead{T$_{\rm eff}$} & \colhead{V$_r$} \\
 & & \colhead{(J2000)} & \colhead{(J2000)} &  &  & \colhead{(arcsec)} & \colhead{(K)} & \colhead{(km s$^{-1}$)}}
\startdata
 & & & & & & & & \\
 E-BSS1&  1090214  &  6.0601164  & $-72.0726528$  &  13.573  &  0.928  &  50.6 & 5013 & $-24.3 \pm 0.05$  \\
 bHB1  &  1109049  &  6.0001040  & $-72.0720222$  &  13.761  &  0.976  &  42.1 & 4896 & $-7.8 \pm 0.06$  \\
 bHB2  &  2625792  &  5.9343037  & $-72.1054776$  &  13.722  &  0.957  & 132.2 & 4940 & $-11.7 \pm 0.06$  \\
\enddata
\tablecomments{Coordinates, $V$ band magnitude, $(V-I)$ color,
  distance from the center, effective temperature and radial velocity
  of the three target stars.  The cluster center used to compute the
  distance from the center is from \citet{miocchi13}.}
\label{tab1}
\end{deluxetable}

\begin{deluxetable}{ccrccccc}
\tablecolumns{5}
\tiny
\tablewidth{0pt}
\tablecaption{Adopted line list}
\tablehead{\colhead{Wavelength} & \colhead{El.} & \colhead{E.P.} & \colhead{log $gf$} & \colhead{EW$_{\rm E-BSS1}$} & \colhead{EW$_{\rm bHB1}$} & \colhead{EW$_{\rm bHB2}$}  \\
\colhead{($\mathring{\rm A}$)} & & \colhead{(eV)} &  & \colhead{(mA)} & \colhead{(mA)} & \colhead{(mA)}} 
\startdata
4834.507 & FeI & 2.42 & $-3.330$ &  29.50 &  39.10 &   0.00 \\
4841.785 & FeI & 4.19 & $-1.880$ &   0.00 &  15.70 &   0.00 \\
4877.606 & FeI & 3.00 & $-3.090$ &   0.00 &  16.20 &   0.00 \\
4885.430 & FeI & 3.88 & $-1.095$ &  63.60 &  66.40 &  61.40 \\
4896.439 & FeI & 3.88 & $-2.020$ &   0.00 &  23.40 &  20.30 \\
4909.383 & FeI & 3.93 & $-1.327$ &  54.60 &  57.70 &  52.90 \\
4917.230 & FeI & 4.19 & $-1.160$ &  51.70 &  52.10 &  47.70 \\
4918.013 & FeI & 4.23 & $-1.340$ &  35.80 &  38.20 &  39.60 \\
4924.770 & FeI & 2.28 & $-2.114$ &  92.90 & 107.50 &  98.70 \\
\enddata
\tablecomments{The entire Table is available in the on-line version. A
  portion is shown here for guidance. Columns 1 to 6 provide:
  wavelength, element, excitation potential, oscillator strength, and
  equivalent width of adopted lines.}  
\label{tab2}
\end{deluxetable}

\begin{landscape}
\begin{deluxetable}{ccccccccccc}
\tablecolumns{11}
\footnotesize
\tablewidth{0pt}
\tablecaption{Abundance ratios of Fe and Ti obtained by  adopting different  stellar mass (gravity) values}
\tablehead{\colhead{Mass} & \colhead{log~$g^{phot}$} & \colhead{$v_{turb}^{spec}$}
& \colhead{[FeI/H]} & \colhead{n(FeI)} & \colhead{[FeII/H]} & \colhead{n(FeII)}
& \colhead{[TiI/H]} & \colhead{n(TiI)} & \colhead{[TiII/H]} & \colhead{n(TiII)} \\
($M_{\odot}$) & (dex) & \colhead{(km s$^{-1}$)} & \colhead{(dex)} &
& \colhead{(dex)} & & \colhead{(dex)} & & \colhead{(dex)} & }
\startdata
 & & & & & E-BSS1 & & & & & \\
\hline
 0.60  &  2.03  &  1.30  &  --0.79 $\pm$ 0.01  &  122  &  --0.97 $\pm$ 0.01  &  9  &  --0.56 $\pm$ 0.01  &  21  &  --0.76 $\pm$ 0.05  &  11  \\
 0.70  &  2.10  &  1.25  &  --0.77 $\pm$ 0.01  &  122  &  --0.92 $\pm$ 0.01  &  9  &  --0.57 $\pm$ 0.01  &  20  &  --0.73 $\pm$ 0.05  &  11  \\
 0.80  &  2.15  &  1.25  &  --0.78 $\pm$ 0.01  &  121  &  --0.90 $\pm$ 0.01  &  9  &  --0.57 $\pm$ 0.01  &  20  &  --0.71 $\pm$ 0.05  &  11  \\
 0.90  &  2.21  &  1.25  &  --0.78 $\pm$ 0.01  &  121  &  --0.87 $\pm$ 0.01  &  9  &  --0.57 $\pm$ 0.01  &  20  &  --0.68 $\pm$ 0.05  &  11  \\
 1.00  &  2.25  &  1.20  &  --0.76 $\pm$ 0.01  &  123  &  --0.84 $\pm$ 0.01  &  9  &  --0.57 $\pm$ 0.01  &  20  &  --0.66 $\pm$ 0.05  &  11  \\
 1.10  &  2.30  &  1.20  &  --0.76 $\pm$ 0.01  &  123  &  --0.82 $\pm$ 0.01  &  9  &  --0.57 $\pm$ 0.01  &  21  &  --0.63 $\pm$ 0.05  &  11  \\
 1.20  &  2.33  &  1.20  &  --0.77 $\pm$ 0.01  &  122  &  --0.80 $\pm$ 0.01  &  9  &  --0.57 $\pm$ 0.01  &  21  &  --0.62 $\pm$ 0.05  &  11  \\
 1.30  &  2.37  &  1.20  &  --0.77 $\pm$ 0.01  &  125  &  --0.79 $\pm$ 0.01  &  9  &  --0.57 $\pm$ 0.01  &  21  &  --0.61 $\pm$ 0.05  &  11  \\
 1.40  &  2.40  &  1.15  &  --0.76 $\pm$ 0.01  &  127  &  --0.76 $\pm$ 0.01  &  9  &  --0.57 $\pm$ 0.01  &  20  &  --0.59 $\pm$ 0.05  &  11  \\
\hline
 & & & & & & & & & & \\
 & & & & & bHB1 & & & & & \\
\hline
 0.60  &  2.06  &  1.35  &  --0.84 $\pm$ 0.01  &  129  &  --0.85 $\pm$ 0.01  & 12  &  --0.66 $\pm$ 0.01  &  26  &  --0.69 $\pm$ 0.05  &  14  \\
 0.70  &  2.13  &  1.35  &  --0.84 $\pm$ 0.01  &  128  &  --0.82 $\pm$ 0.01  & 12  &  --0.66 $\pm$ 0.01  &  26  &  --0.66 $\pm$ 0.05  &  14  \\
 0.80  &  2.18  &  1.35  &  --0.84 $\pm$ 0.01  &  128  &  --0.80 $\pm$ 0.01  & 12  &  --0.66 $\pm$ 0.01  &  26  &  --0.64 $\pm$ 0.05  &  14  \\
 0.90  &  2.23  &  1.35  &  --0.84 $\pm$ 0.01  &  127  &  --0.78 $\pm$ 0.01  & 12  &  --0.66 $\pm$ 0.01  &  26  &  --0.62 $\pm$ 0.05  &  14  \\
 1.00  &  2.28  &  1.35  &  --0.84 $\pm$ 0.01  &  126  &  --0.75 $\pm$ 0.01  & 12  &  --0.66 $\pm$ 0.01  &  26  &  --0.59 $\pm$ 0.05  &  14  \\
 1.10  &  2.32  &  1.35  &  --0.84 $\pm$ 0.01  &  126  &  --0.73 $\pm$ 0.01  & 12  &  --0.66 $\pm$ 0.01  &  26  &  --0.58 $\pm$ 0.05  &  14  \\
 1.20  &  2.36  &  1.35  &  --0.84 $\pm$ 0.01  &  126  &  --0.72 $\pm$ 0.01  & 12  &  --0.67 $\pm$ 0.01  &  26  &  --0.57 $\pm$ 0.05  &  14  \\
 1.30  &  2.39  &  1.30  &  --0.83 $\pm$ 0.01  &  126  &  --0.69 $\pm$ 0.01  & 12  &  --0.67 $\pm$ 0.01  &  26  &  --0.55 $\pm$ 0.05  &  14  \\
 1.40  &  2.43  &  1.30  &  --0.83 $\pm$ 0.01  &  127  &  --0.67 $\pm$ 0.01  & 12  &  --0.67 $\pm$ 0.01  &  26  &  --0.54 $\pm$ 0.05  &  14  \\
\hline
 & & & & & & & & & & \\
 & & & & & bHB2 & & & & & \\
\hline
 0.60  &  2.06  &  1.20  &  --0.81 $\pm$ 0.01  &  124  &  --0.81 $\pm$ 0.02  & 10  &  --0.68 $\pm$ 0.01  &  22  &  --0.67 $\pm$ 0.05  &  13  \\
 0.70  &  2.13  &  1.20  &  --0.81 $\pm$ 0.01  &  128  &  --0.78 $\pm$ 0.02  & 10  &  --0.68 $\pm$ 0.01  &  22  &  --0.64 $\pm$ 0.05  &  13  \\
 0.80  &  2.19  &  1.20  &  --0.82 $\pm$ 0.01  &  129  &  --0.75 $\pm$ 0.02  & 10  &  --0.68 $\pm$ 0.01  &  22  &  --0.61 $\pm$ 0.05  &  13  \\
 0.90  &  2.24  &  1.15  &  --0.80 $\pm$ 0.01  &  128  &  --0.72 $\pm$ 0.02  & 11  &  --0.68 $\pm$ 0.01  &  22  &  --0.59 $\pm$ 0.05  &  13  \\
 1.00  &  2.29  &  1.15  &  --0.78 $\pm$ 0.01  &  127  &  --0.68 $\pm$ 0.02  & 11  &  --0.68 $\pm$ 0.01  &  22  &  --0.57 $\pm$ 0.05  &  13  \\
 1.10  &  2.33  &  1.15  &  --0.80 $\pm$ 0.01  &  125  &  --0.68 $\pm$ 0.02  & 11  &  --0.68 $\pm$ 0.01  &  22  &  --0.56 $\pm$ 0.05  &  13  \\
 1.20  &  2.37  &  1.10  &  --0.79 $\pm$ 0.01  &  125  &  --0.65 $\pm$ 0.02  & 11  &  --0.68 $\pm$ 0.01  &  22  &  --0.54 $\pm$ 0.05  &  13  \\
 1.30  &  2.40  &  1.10  &  --0.78 $\pm$ 0.01  &  124  &  --0.63 $\pm$ 0.02  & 11  &  --0.69 $\pm$ 0.01  &  22  &  --0.53 $\pm$ 0.05  &  13  \\
 1.40  &  2.43  &  1.10  &  --0.79 $\pm$ 0.01  &  124  &  --0.62 $\pm$ 0.02  & 11  &  --0.69 $\pm$ 0.01  &  22  &  --0.52 $\pm$ 0.05  &  13  \\
\enddata
\tablecomments{
Iron and titanium abundance ratios obtained for the program stars by
adopting the mass values listed in the first column.  During the
analysis the effective temperature (listed in Table \ref{tab1}) and
surface gravity (column 2) have been kept fixed, while the
microturbulent velocity (column 3) has been spectroscopically
optimized. The abundances obtained from neutral and ionized lines, and
the number of lines used are listed in columns 4--11 (see labels). We
adopted the solar reference values of \cite{grevesse98}.}
\label{tab3}
\end{deluxetable}
\end{landscape}

\end{document}